\documentclass[prl,aps,twocolumn,superscriptaddress,showpacs]{revtex4}
\usepackage{latexsym}
\usepackage{graphicx}

\newcommand{\be}{\begin{equation}}
\newcommand{\ee}{\end{equation}}
\newcommand{\ba}{\begin{eqnarray}}
\newcommand{\ea}{\end{eqnarray}}
\newcommand{\baa}{\begin{eqnarray*}}
\newcommand{\eaa}{\end{eqnarray*}}

\def\be{\begin{equation}}
\def\ee{\end{equation}}
\def\bea{\begin{eqnarray}}
\def\eea{\end{eqnarray}}

\def\C60{A$_x$C$_{60}$}

\def\HgCu3{HgCa$_2$Cu$_3$O$_{8+y}$}
\def\HgCu4{HgBa$_2$Ca$_3$Cu$_4$O$_{10+y}$}
\def\TlCu{Tl$_2$Ba$_2$CuO$_{6+\delta}$}
\def\TlCu3{Tl$_2$Ba$_2$Ca$_2$Cu$_3$O$_{10+y}$}
\def\TlCu4{Tl$_2$Ba$_2$Ca$_3$Cu$_4$O$_{12+y}$}

\def\BiCu3{Bi$_2$Sr$_2$Ca$_{2}$Cu$_3$O$_y$}

\def\8LSCO{La$_{1.88}$Sr$_{.12}$CuO$_4$}
\def\110LNSCO{La$_{1.5}$Nd$_{0.4}$Sr$_{0.1}$CuO$_{4}$}
\def\stage4LCO{La$_{2}$CuO$_{4+\delta}$}
\def\Y248{YBa$_2$Cu$_4$O$_8$}

\def\NbSe2{NbSe$_2$}
\def\TaSe2{TaSe$_2$}
\def\TiSe2{TiSe$_2$}

\begin{document}

\title{Block Antiferromagnetism and Checkerboard Charge Ordering in Alkali-doped Iron Selenides $R$$_{1-x}$Fe$_{2-y}$Se$_2$}

\author{Wei Li}
\affiliation{Department of Physics, Fudan University, Shanghai 200433, China}
\author{Shuai Dong}
\affiliation{Department of Physics, Southeast University, Nanjing 211189, China}
\author{Chen Fang}
\affiliation{Department of Physics, Purdue University, West Lafayette, Indiana 47907, USA}
\author{Jiangping Hu}
\email{jphu@iphy.ac.cn}
\affiliation{Beijing National Laboratory for Condensed Matter Physics, Institute of Physics,
Chinese Academy of Sciences, Beijing 100080, China}
\affiliation{Department of Physics, Purdue University, West Lafayette, Indiana 47907, USA}

\date{\today}


\begin{abstract}
By performing first-principles electronic structure calculations and analyzing effective magnetic model of alkali-doped iron selenides, we show that the materials without iron vacancies should approach a novel checkerboard phase in which each four Fe sites group together in tetragonal structure. The checkerboard phase is the ground state with a block antiferromagnetic (AFM) order and a small charge density wave order in the absence of superconductivity. Both of them can also coexist with superconductivity. The results explain mysterious $2\times 2$ ordered patterns and hidden orders observed in various different experiments, clarify the missing link between AFM and superconducting phases, suggest that the block-AFM state is the parent state, and unify the understanding of various observed phases in alkali-doped iron selenides.
\end{abstract}
\maketitle
The newly discovered alkali-doped iron selenide superconductors\cite{Guo2010,Fang2010,Liu2011} have attracted much research attention because of several distinct characters that are noticeably absent in other iron-based superconductors, such as the absence of hole pockets at $\Gamma$ point of Brillouin zone in their superconducting (SC) phases\cite{ZhangY2010,WangXP2011,Mou2011} and AFM ordered insulating phases\cite{wbao1,wbao2,mwang} with very high N\'{e}el transition temperatures in their parental compounds\cite{Liu2011}. Due to these distinct physical characters from their pnictide counterparts, $R_{1-x}$Fe$_{2-y}$Se$_2$ are expected to be ideal grounds to test theoretical models of iron-based superconductors. Models based on different mechanisms have suggested different pairing symmetries for $R_{1-x}$Fe$_{2-y}$Se$_2$: weak coupling approaches based on spin-excitation mediated pairing predict a $d$-wave pairing symmetry\cite{cth,maitikorsh,Youy2011,Maier2011}, strong coupling approaches\cite{Fang2011,huj,Yu2011} which emphasize the importance of next nearest neighbor (NNN) AFM local exchange coupling suggest the pairing symmetry is a robust $s$-wave, not different from the $S^\pm$-wave symmetry obtained in their pnictide counterparts and models with orbital fluctuation mediated pairing suggest a $S^{++}$-wave pairing for both iron selenide and pnictide materials\cite{saito}.

While the iron-selenide superconductors have generated considerable excitement, there are deep confusions regarding the delicate interplay between Fe vacancies, magnetism and superconductivity. Many latest experimental results in $R_{1-x}$Fe$_{2-y}$Se$_2$ indicate that the insulating AFM and SC phases are phase separated\cite{hhwenps,xhps, dlfengps, yywangps, chenxps,wangnlps}. In particular, the recent scanning tunneling microscopy (STM) measurements on K$_{1-x}$Fe$_{2-y}$Se$_2$ clearly suggest phase separation\cite{chenxps,yywangps}. The material was shown to be phase separated into iron vacancy ordered regions and iron vacancy free regions. The former is insulating and shows a $\sqrt{5}\times \sqrt{5}$ vacancy ordered pattern while the latter is SC.
The neutron scattering experiments have shown that the $\sqrt{5}\times \sqrt{5}$ vacancy ordered phase is also AFM ordered\cite{wbao1}. More surprisingly, the AFM order is affected by SC pairing\cite{wbao2}, a result difficult to be understood within the picture of phase separation. Besides the $\sqrt{5}\times \sqrt{5}$ vacancy ordered phase, an additional insulating phase with a $2\times 2$ ordered pattern was also observed\cite{xhps,dlfengps,zwang}. Moreover, in the SC state where there is few vacancy, both STM\cite{yywangps,chenxps} and angle resolved photoemission spectroscopy (ARPES) experiments\cite{Mou2011} suggest that there is an additional symmetry breaking order in the SC phases\cite{chenxps,yywangps,Mou2011}. The microscopic origin of this order and how it is related to the AFM phase are not understood.

In this Letter, we show that alkali-doped iron selenide superconductors without iron vacancies should approach a checkerboard phase in which each four Fe sites group together in a tetragonal structure. This broken symmetry state is essentially driven by the same magnetic exchange couplings that drive the insulating AFM phase in the $\sqrt{5}\times \sqrt{5}$ vacancy ordered state. We perform first-principles electronic structure calculations and develop an effective magnetic model to show the existence of such a broken symmetry state. The checkerboard phase is the ground state with a block-AFM (BAF) order in the absence of superconductivity. The BAF fluctuations and the checkerboard lattice distortion are strongly coupled. A weak BAF order and the checkerboard lattice distortion can coexist with superconductivity. These results essentially suggest the BAF ordered state are the parent state of alkali-doped iron selenide superconductors. The results consistently explain the mysterious $2\times 2$ ordered pattern which was misunderstood as another vacancy ordered state and the STM and ARPES experimental results\cite{yywangps, Mou2011}. This study clarifies the missing link between AFM and SC phases and essentially unifies the understanding of various observed phases\cite{xhps}.

\begin{figure}
\begin{center}
\tabcolsep=0cm
\begin{tabular}{c}
\includegraphics[width=8cm,height=4.5cm]{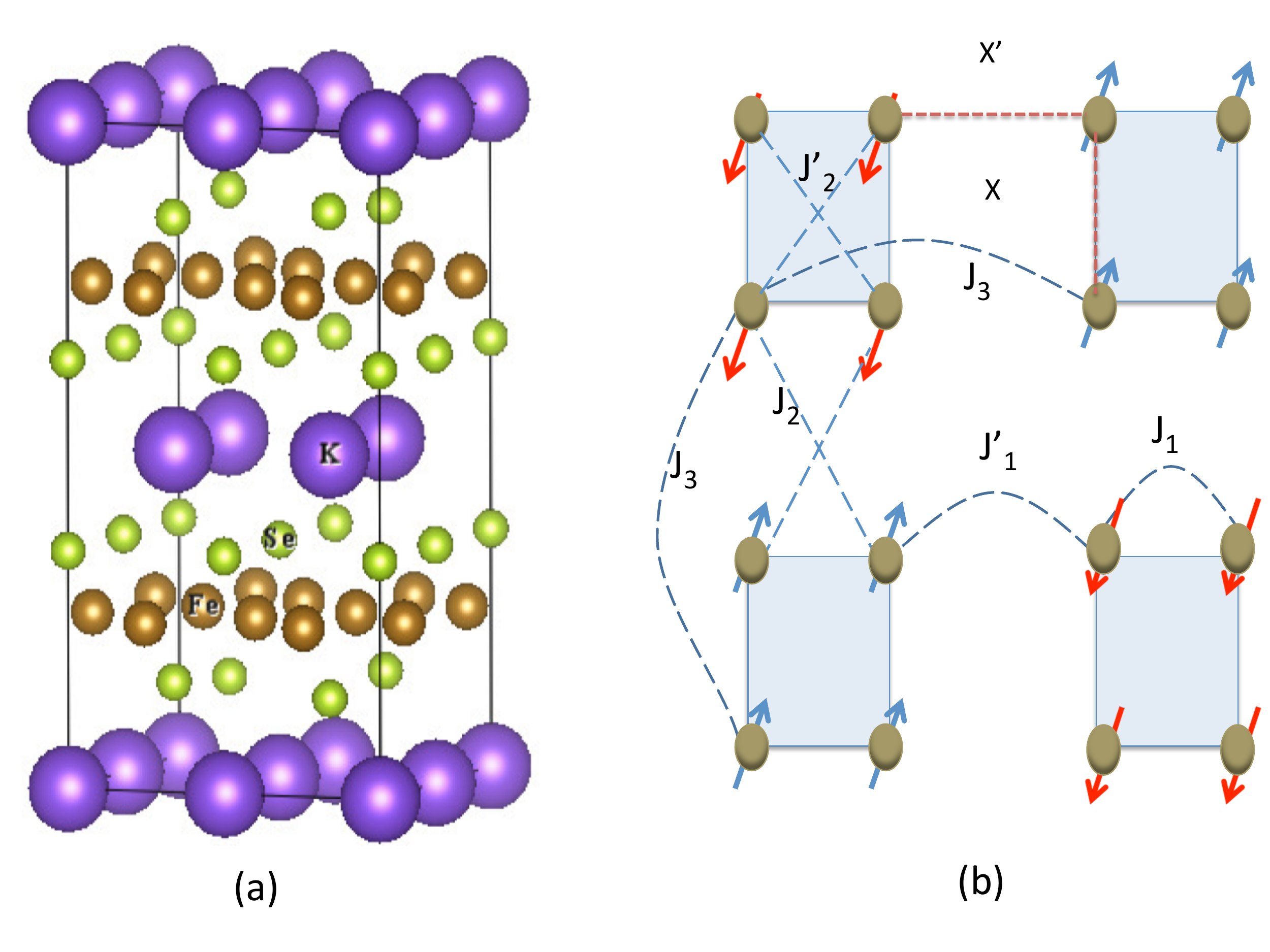}
\end{tabular}
\end{center}
\caption{(Color online) (a) The crystal structure of  KFe$_{2}$Se$_{2}$ compound in our calculation. The system consists of K (royal), Fe (dark yellow) and Se (green) atoms; (b) Schematic checkerboard lattice structure and spin ordering pattern in the BAF state. The lattice distortion is labeled by lattice constants $X$ and $X'$. The magnetic exchange couplings are also indicated.}\label{fig:fig1}
\end{figure}

{\it Results from LDA calculations} We start with the following simple question: if the system is free of iron vacancies, what should be the ground state if it does not become SC? To answer this question, we perform the first principles calculation to investigate the ground state of an iron vacancy free domain. We calculated the energy of a number of different possible magnetically ordered states, including non-magnetic (NM), ferromagnetic (FM), collinear-AFM (CAF, the state observed in iron-pnictides\cite{Zhaoj2008,Zhaoj2009}), bicollinear-AFM (BCAF, the state observed in FeTe)\cite{Bao2008,Lis2008a,Ma2009a,Fang2009b} and BAF whose pattern is shown in Fig. \ref{fig:fig1}(b) where four Fe sites group to form a super cell. All these calculations were performed using the projected augmented wave method\cite{Ref10} as implemented in the VASP code\cite{Ref11}, and the Perdew-Burke-Ernzerhof (PBE) exchange correlation potential\cite{Ref12} was used. A 500eV cutoff in the plane wave expansion ensures the calculations converge to $10^{-5}$ eV. For the BAF state, all atomic positions and the lattice constants were optimized until the largest force on any atom was 0.005eV/\AA. We used a $9\times 9\times 5$ Monkhorst-Pack k-grid Brillouin zone sampling throughout all of calculations.

The ground state energies, the magnetically ordered moment and their lattice constants in various states are listed in Table \ref{table}. The BAF state clearly has the lowest energy. Without turning on $U$, our LDA results show that the BAF ordered state is metallic. However if a finite $U>1.0$eV is added, the LDA+$U$ calculation shows that the state becomes insulating. There are strong lattice distortion in the BAF state. The lattice constant $X$ (between two nearest sites in one supercell) and $X'$ (between two supercells) as labeled in Fig. \ref{fig:fig1}(b) are $X$=2.6\AA, $X'$=2.85\AA\ for $U=0$eV and $X$=2.59\AA, $X'$=3.03\AA\ for $U=2$eV. This lattice distortion is comparable to the lattice distortion in the $\sqrt{5}\times\sqrt{5}$ vacancy ordered K$_{0.8}$Fe$_{1.6}$Se$_{2}$ phase\cite{wbao1}.
\begin{table}
\caption{Geometric, energetic and magnetic properties of KFe$_{2}$Se$_{2}$. Results in the NM/FM/AFM/CAF/BCAF/BAF and optimized BAF configurations using fully optimized structures are all shown. $\Delta E$ is the total energy difference per iron atom in reference to
the unoptimized experimental structure\cite{Guo2010}, and $m_{Fe}$ is the local magnetic moment on Fe.}\label{table}
\begin{ruledtabular}
\begin{tabular}{ccccc}
KFe$_{2}$Se$_{2}$  & $\Delta$E (eV/Fe) & a(\AA) & c(\AA) &  $m_{Fe}$($\mu_{B}$)    \\
\hline
  NM   &  0         &  3.9136  & 14.0367 & 0   \\
  FM   &  -0.2400   &  3.9136  & 14.0367 & 2.781   \\
  AFM  & -0.2384   &  3.9136  & 14.0367 & 2.135  \\
  CAF  &-0.3510&  3.9136  & 14.0367 & 2.446  \\
  BCAF &-0.3159&  3.9136  & 14.0367 &2.556\\
  BAF  &       -0.3127&  3.9136  & 14.0367 &2.552 \\
  \hline
  BAF(opt.)  & -0.3568&  3.8553  & 14.4099 &2.635
%
\end{tabular}
\end{ruledtabular}
\end{table}


From Table \ref{table}, the energies of the BCAF state and the BAF state are almost degenerate in the unoptimized lattice tetragonal structure. The actual ground state, in fact, depends on the optimization of lattice structure. In the previous first principle calculations, the BCAF state was shown to be the lowest energy state in the optimized lattice which has monoclinic distortion\cite{Yan2011,Yan2011a}. The monoclinic distortion and the BCAF order can be strongly coupled with each other because they break the same lattice rotational symmetry. Such a strong coupling was observed in FeTe in which a single strong first order phase transition where both the BCAF ordering and the tetragonal-monoclinic distortion take place\cite{Bao2008}. However, in KFe$_2$Se$_2$, no monoclinic lattice distortion has been observed. Without allowing the monoclinic lattice distortion, the BAF state becomes the ground state.

\begin{figure}
\includegraphics[width=6cm, height=5cm]{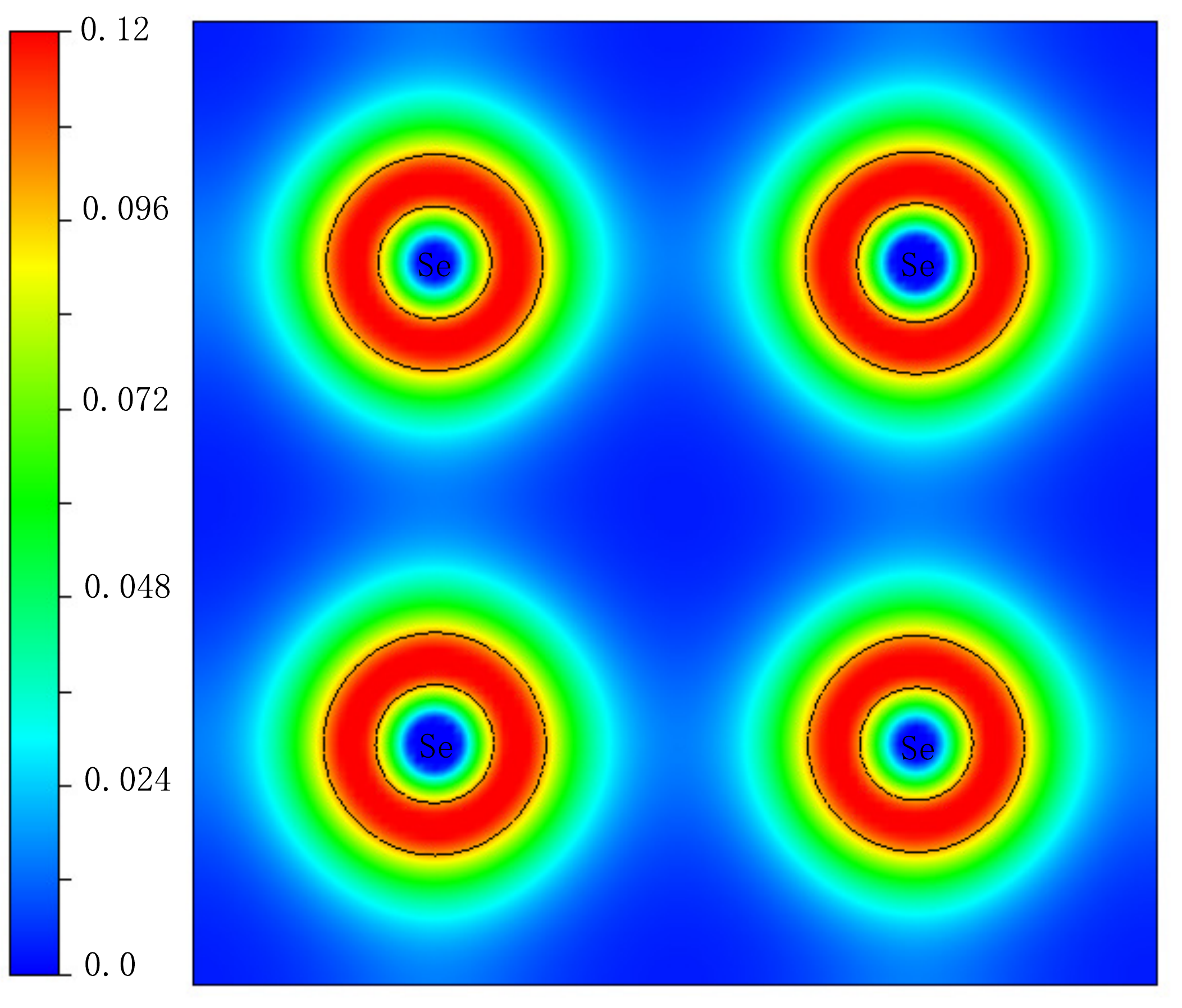}
\caption{(Color online) Charge density distribution ($e/bohr^3$)
in the ($001$) plane crossing the first Se atoms layer with BAF
order state within the LDA+$U$ ($U=2$eV) calculations.}
\label{fig:fig2}
\end{figure}

From the similar symmetry analysis, the BAF state must strongly couple to the lattice distortion shown in Fig. \ref{fig:fig1}(b). The lattice distortion quadroples the lattice unit cell to form a checkerboard pattern. In such a checkerboard lattice, the charge ordering can take place. If we calculate the electron density distribution around Se atoms in the Se-layer above the iron-layer, a charge ordering on the Se-layer is observed as shown in Fig. \ref{fig:fig2}, which was observed in recent STM experiment\cite{yywangps}.

\begin{figure}
\begin{center}
\tabcolsep=-0.2cm
\begin{tabular}{c}
\includegraphics[width=8cm,height=6.2cm]{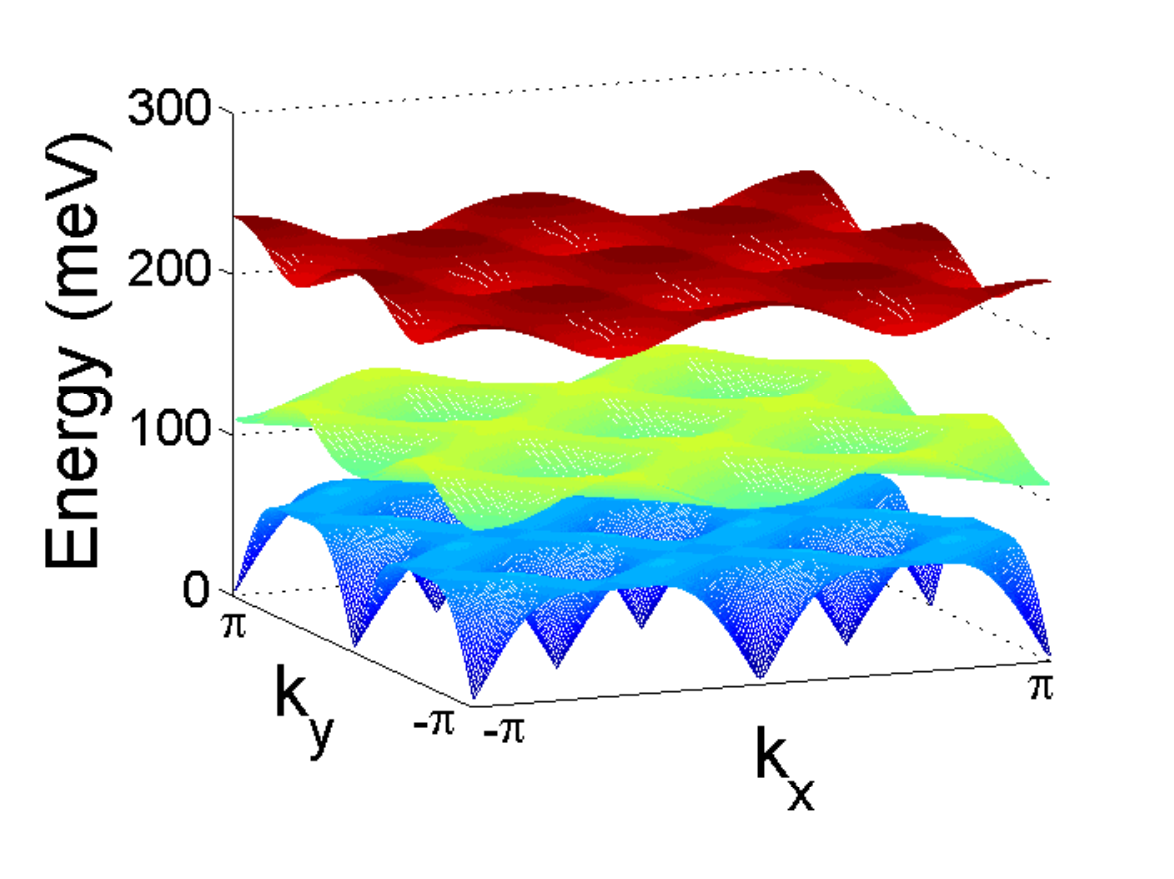} \\
\includegraphics[width=8cm,height=6.2cm]{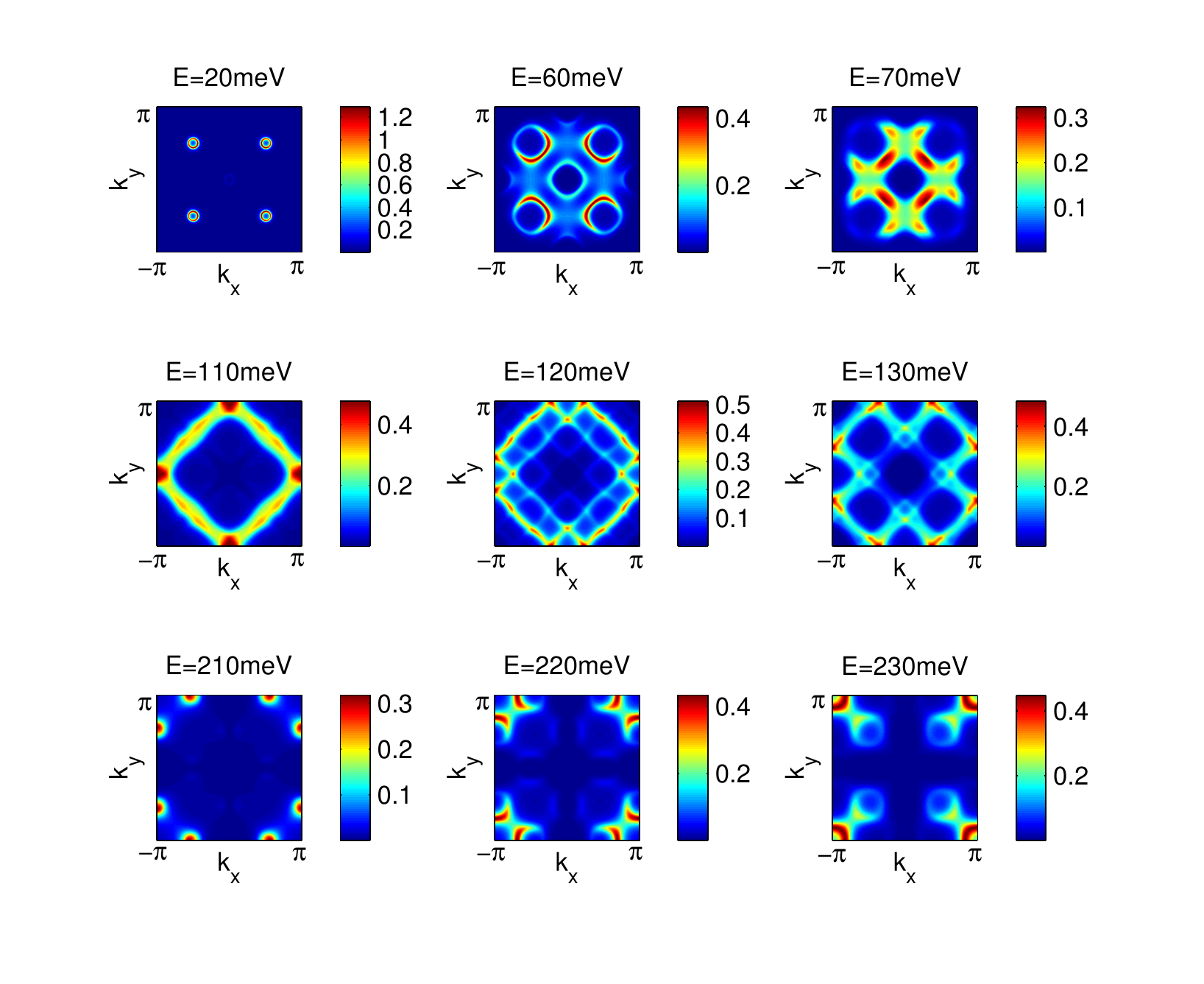}
\end{tabular}
\end{center}
\caption{(Color online) Spin wave dispersion (up) and imaginary part of dynamic
susceptibility (down) for the BAF state with $J_1$-$J'_1$-$J_2$-$J_3$ model at $J_1=-36$meV, $J'_1=15$meV, $J_2=14$meV,
$J_3=9$meV, taken from Ref. \onlinecite{mwang}. The profile
of the imaginary part of the dynamic susceptibility is plotted at various
energies with an energy resolution of 5meV, and it is given in arbitrary
unit.}\label{intensity}
\end{figure}

{\it Magnetic models} Now we discuss the effective magnetic model that can interpret above calculation results. It has been shown a magnetic exchange $J_1$-$J_2$-$J_3$-$K$ model\cite{Hu2011b} where $J_1, J_2$ and $J_3$ are the nearest neighbor (NN), NNN, and the next NNN (NNNN) exchange couplings respectively, and $K$ is a spin biquadratic coupling term between two nearest neighbor sites is a good approximation to describe iron-chalcogenides when the lattice distortion is ignored\cite{Ma2009a,Fang2009b,Hu2011a,Hu2011b}. In Table \ref{table}, the energies of the BCAF and BAF states in the unoptimized lattice tetragonal structure are almost degenerate. This degeneracy is a strong support of the model since these two states are exactly degenerate in the $J_1$-$J_2$-$J_3$-$K$ classical spin model\cite{Hu2011b}.  In the magnetically ordered state,   the lattice distortion takes place and the tetragonal symmetry is broken. The nearest neighbor exchange coupling $J_1$ can take two different values $J_1$ and $J'_1$ as shown in Fig. \ref{fig:fig1}(b). The biquadratic coupling $K$  can be decoupled and treated as an effective difference between $J_1$ and $J'_1$ as well\cite{Hu2011b,wy}. In general, the NNN $J_2$ can also take two different values, $J_2$ and $J'_2$ as also shown in Fig. \ref{fig:fig1}(b). However, as being proved in other iron-based superconductors, the NNN coupling $J_2$ is rather robust against lattice distortion. The difference between $J_2$ and $J'_2$ is rather small. Therefore, the effective magnetic exchange model in magnetically ordered state is given by $J_1$-$J'_1$-$J_2$-$J_3$ with $J_1$ being strongly FM and $J_{2,3}$ both being AFM. $J'_1$ can be weak FM or weak AFM. The similar model has been shown to describe the magnetism of the $\sqrt{5}\times\sqrt{5}$ vacancy ordered K$_{0.8}$Fe$_{1.6}$Se$_{2}$ phase\cite{Hu2011a,mwang}. Therefore, while the exact values of the magnetic exchange couplings can not be accurately obtained from LDA calculations, since the lattice distortions in both cases are similar, it is reasonable to believe that these values should not be too different from those of K$_{0.8}$Fe$_{1.6}$Se$_{2}$ which has been measured by fitting neutron scattering experiments\cite{mwang}. The measured values, which are specified in Fig. \ref{intensity}, also give the BAF order ground state. The saved energy from magnetic exchange coupling per site is given by $(-J_1+2J_3+J'_1)S^2$. This energy is sightly smaller than the saved magnetic energy in the vacancy ordered K$_{0.8}$Fe$_{1.6}$Se$_{2}$\cite{Hu2011a,mwang}. The spin wave dispersion and the imaginary part of dynamic spin susceptibility of the BAF phase are shown in Fig. \ref{intensity}.  

\begin{figure}
\includegraphics[width=9cm,height=6cm]{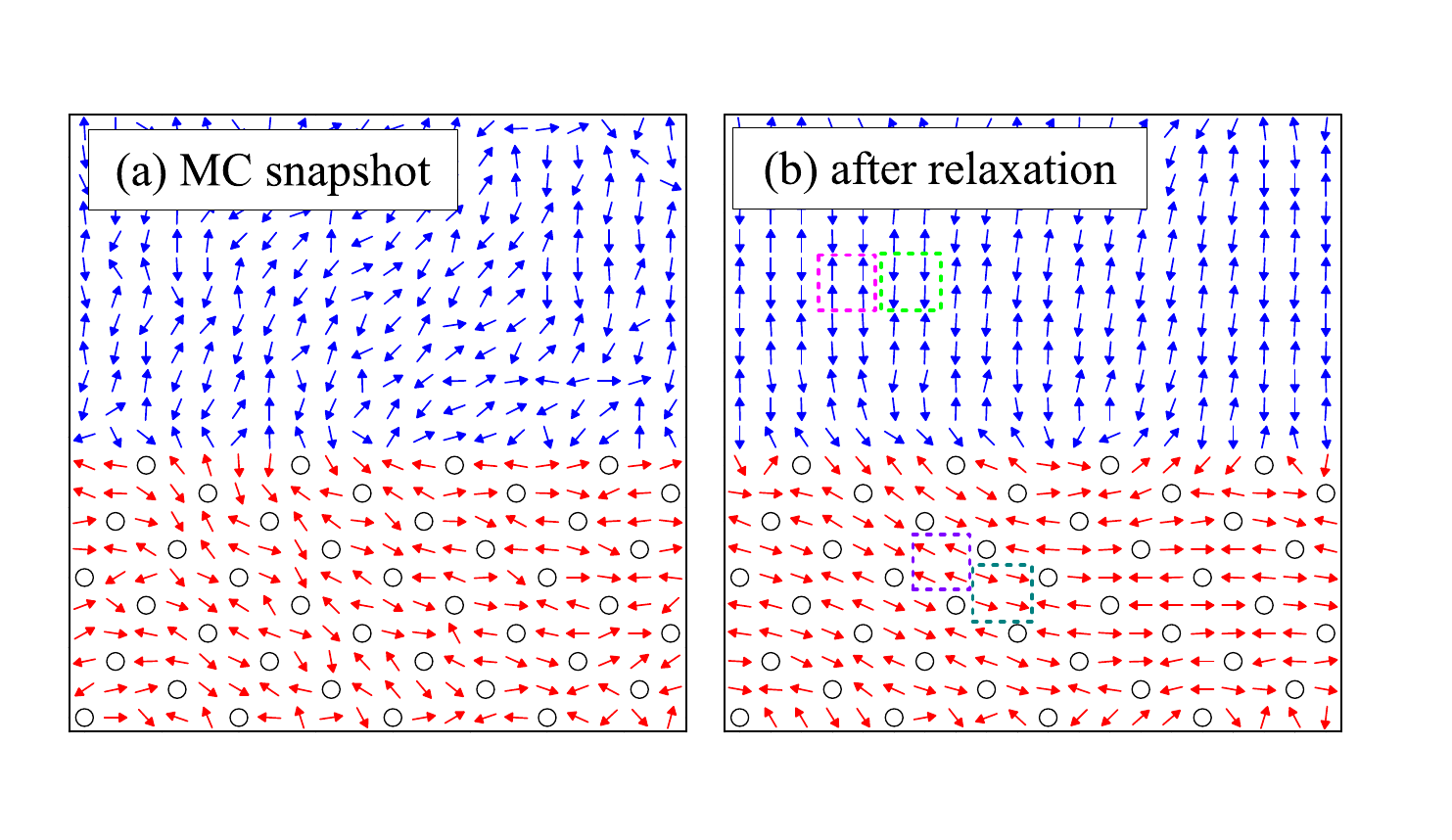}
\caption{(Color online) (a) A typical MC snapshot (after $3\times 10^4$ MC steps) of the classical $J_1$-$J_2$-$J_3$-$K$ magnetic model. (b) The spin pattern after the zero-$T$ relaxation. In (a) and (b), spins in regions without vacancies and with vacancies are in blue and red, respectively. Black circles denote the vacancies.}
\label{mc}
\end{figure}

There is another interesting prediction if the same magnetic model describes both K$_{0.8}$Fe$_{1.6}$Se$_{2}$ and K$_x$Fe$_2$Se$_2$. As mentioned before, K$_{0.8}$Fe$_{1.6}$Se$_{2}$ and K$_x$Fe$_2$Se$_2$ are two phase separated regions. If the same effective magnetic model describes both structures, it is very interesting to inquire into the magnetic configurations near the boundary of these two structures. We perform a Monte Carlo (MC) simulation on the $J_1$-$J_2$-$J_3$-$K$ model\cite{Hu2011b} to address this problem. A simple numerical simulation, which includes a standard Markov Chain MC simulation followed by a zero-temperature relaxation process, is performed to qualitatively investigate the magnetic orders near phase boundaries. A two-dimensional spin lattice [$L_x\times(L_{y1}+L_{y2})$] is used with periodic boundary conditions (PBCs). Vacancies with the $\sqrt{5}\times\sqrt{5}$-pattern is created in $L_{y1}$ regions for the K$_{0.8}$Fe$_{1.6}$Se$_{2}$ phase. A general result we obtained as shown in Fig. \ref{mc} is that the spin directions between $L_{y1}$ and $L_{y2}$ regions are noncollinear. Since experimentally, the ordered AFM moment is along c-axis in K$_{0.8}$Fe$_{1.6}$Se$_{2}$\cite{wbao1}, this result suggests that the ordered moment in the BAF state must be in the plane. This non-collinearity stems from the presence of vacancies and intrinsic magnetic frustration among the magnetic exchange couplings, similar to the study in the frustrated $J_1$-$J_2$ model\cite{Hu2011c}. Recent STM results have provided an evidence supporting this prediction.
It was shown that the magnetic moment induced by individual vacancy in the SC state is indeed in the plane\cite{chenxps}.

The phase separation between the vacancy ordered BAF state and the SC state has blurred the interplay between magnetism and superconductivity in alkali-doped iron selenide. The above results also clarify the connection. In iron-pnictides, as increasing doping suppresses CAF order, superconductivity develops. The magnetic order is able to coexist with superconductivity in a small doping region\cite{johnston}. Even in the region where the magnetic order is completely suppressed, orthorhombic lattice distortion which couples the fluctuating short range CAF order\cite{Fang2008d,xu}, can survive and coexist with superconductivity. Our result suggests that the similar physics can take place in $R$$_{1-x}$Fe$_{2-y}$Se$_2$. The absence of iron vacancies in the SC state suggests that the true SC material has a chemical formula $R_{1-x}$Fe$_2$Se$_2$. The parent state of this material should be a BAF state. Increasing doping suppresses the BAF state and leads to SC. While it is still difficult to determine whether the BAF and SC can coexist, we can safely argue that, similar to iron-pnictides, a lattice distortion as shown in Fig. \ref{fig:fig1}(b) that couples to the short range BAF fluctuation should be able to coexist with SC.

This picture provide explanations to many puzzling phenomena observed in alkali-doped iron selenides $R_{1-x}$Fe$_{2-y}$Se$_2$. First, in ARPES measurements, a weak but large electron pocket at $\Gamma$ point was observed\cite{Mou2011}. This pocket is almost identical to the electron pockets at $M$ point, suggesting the electron pocket is a folded pocket due to translational symmetry breaking in the SC state. Moreover, in recent STM experiment\cite{yywangps}, a $2\times2$ charge density modulation with respect to Fe-lattice was observed to coexist with the SC phase. These electronic superstructure are consistent with the checkerboard phase. Second, neutron scattering experiments suggested that the vacancy ordered-AFM state interacts strongly with superconductivity\cite{wbao2}. In a phase separation scenario, such a strong interaction is hard to understood. Our results resolve such a dilemma. The experiment can be easily understood because the BAF state strongly interacts with both the vacancy ordered-AFM state and the SC state. The development of superconductivity is expected to strongly suppress the BAF state. Finally, the $2\times 2$ ordering in insulating samples observed by transmission electron microscopy (TEM)\cite{xhps,zwang} can be naturally interpreted as our checkerboard state with the BAF order. Previously it was interpreted as another vacancy ordered phase\cite{xhps, zwang}. Such an interpretation is unlikely because the TEM signal of this order is much weaker than that of the $\sqrt{5} \times \sqrt{5}$ vacancy order.

In summary, we show that alkali-doped iron selenides $R$$_{1-x}$Fe$_{2-y}$Se$_2$ has a checkerboard phase in which each four Fe sites group together in a tetragonal structure. The checkerboard phase approaches a BAF order in the absence of superconductivity. The phase also exhibits small charge density modulation on Se sites.  Magnetic properties related to this state are calculated. Combining with the strong experimental evidence of phase separation between vacancy ordered and vacancy free phases, we suggest the checkerboard phase is the parent state of the superconductor.

{\it Acknowledgement:} We thank H. Ding, D. L. Feng, P. C. Dai,
N. L. Wang, H. H. Wen, X. Chen, Q. K. Xue, T. Xiang and Y. Y. Wang for useful discussion. S.D. was supported by the 973 Projects of China (2011CB922101), NSFC (11004027) and NCET (10-0325)

\end{document}